\documentclass[aps,prd,preprint,groupedaddress]{revtex4-1}

 \usepackage{graphicx} 
\usepackage{color}
 \usepackage{soul} 
 
\begin{document}

\title{  Holographic model for heavy vector meson masses    }

\author{Nelson R. F. Braga}\email{braga@if.ufrj.br}
\affiliation{Instituto de F\'{\i}sica,
Universidade Federal do Rio de Janeiro, Caixa Postal 68528, RJ
21941-972 -- Brazil}

\author{ M. A. Martin Contreras}\email{ma.martin41@uniandes.edu.co}
\affiliation{ High Energy Group, Department of Physics, Universidad de los Andes, Carrera 1, No 18A - 10, Bloque Ip, ZIP 111711, Bogot\'{a}, Colombia}

\author{Saulo Diles}
\email{smdiles@if.ufrj.br}
\affiliation{Instituto de F\'{\i}sica,
Universidade Federal do Rio de Janeiro, Caixa Postal 68528, RJ
21941-972 -- Brazil}


\begin{abstract}
The experimentally observed spectra of heavy vector meson radial excitations show  a dependence 
on two different energy parameters. 
One is associated with the quark mass  and the other with the binding energy levels of the quark anti-quark pair.  The first is present in the large mass of the first state while the other corresponds to the small mass splittings between   radial excitations.  In this article we show how to reproduce such a behavior with reasonable precision 
using a holographic model.  In the dual picture, the large energy scale shows up from a   bulk mass and the small  scale comes from the position of anti-de Sitter (AdS)  space where field correlators are calculated.    The model determines  the masses of  four  observed S-wave states of charmonium and  six  
S-wave states of  bottomonium with  $ \, 6.1 \%$    rms error.  
In consistency with the physical picture, the large energy parameter is flavor dependent, while the 
small parameter, associated with quark anti-quark interaction is the same for charmonium and bottomonium
states.    
 
\end{abstract}


\maketitle

\section{ Introduction }   

Vector mesons made of light quarks present an approximately linear relation between the mass squared
and the radial excitation number $n$, the so called radial Regge trajectory:  $ m_n^ 2 \sim \alpha n $. So, the mass spectrum can be approximately described by just one dimensionfull parameter (energy scale), related to the interaction between the quark and the anti-quark. 

For heavy vector mesons  the situation is different, as one can see from the experimental values of the masses of the $n= 1,2,3,4 ,$  S-wave states of charmonium \cite{Agashe:2014kda}  $ m_1 = 3097 $ MeV, $ m_2  = 3686 $ MeV , $ m_3  = 4039 $ MeV, $ m_4  =  4441 $ MeV. The mass gaps  $ (m_{n+1} -  m_n)  $  are much smaller than the mass  of the first state. 
This suggests the existence of two different energy scales. A large one, associated with the heavy constituent  quark masses and a smaller one,  that appears in the gaps between radial excitations, related to the energy levels of quark anti-quark interaction (binding energy).  A similar behavior is observed for bottomonium states,
where the difference in the scales is even more evident. The Regge trajectories for heavy quarkonium are discussed, for example, in \cite{Gershtein:2006ng}.

AdS/QCD models, like the hard wall \cite{Polchinski:2001tt,BoschiFilho:2002ta,BoschiFilho:2002vd} ,
are motivated by gauge string duality \cite{Maldacena:1997re,Gubser:1998bc,Witten:1998qj} and provide nice descriptions of mass spectra of glueballs and vector mesons made of light
quarks.  
A recent review of hard wall and other holographic models developed afterward 
can be found in  \cite{Brodsky:2014yha}. 

Heavy vector mesons have been discussed in the context of AdS/QCD models in refs. 
\cite{Hong:2003jm,Kim:2007rt,Fujita:2009wc,Fujita:2009ca,Grigoryan:2010pj,Branz:2010ub,Gutsche:2012ez,
Afonin:2013npa,Hashimoto:2014jua}.  
However, concerning the mass spectra of S-wave states of heavy quarkonium,  there is no accurate predictive AdS/QCD model available in the literature. 
The previous studies either provide masses with  large errors with respect to experimentally observed data, or depend on many parameters, thus lacking of predictivity. 
In particular, the simplest picture that one could draw for the heavy quarkonium states is that the mass spectrum should depend on the quark mass, that is flavor dependent, and on the quark anti-quark interaction, that is flavor independent.  This simple physical picture is absent in these previous works and will emerge in the present article. 

 We present here a holographic AdS/QCD model that describes the masses of the S-wave states of charmonium and bottomonium with just 3 parameters that have a very clear physical interpretation:  one is associated with the mass of $c$  quark, the other with the mass of $b$  quark and the third  with  the flavor independent quark interaction.   
 The model reproduces the masses of ten states of charmonium and bottomonium with good precision, 
  characterized by  $ 6.1 \% $ rms error.

\section{ Holographic picture of heavy vector mesons }

The current $ J^\mu = \bar{q}\gamma^\mu q \,$ associated with a heavy vector meson is assumed to be dual to the $V_\mu$ components of a massive vector field  $V_m = (V_\mu,V_z)\,$ ($\mu = 0,1,2,3$) living in anti-de sitter space:
\begin{equation}
 ds^2 =     \frac{R^2}{z^ 2 }(-dt^2 + d\vec{x}\cdot d\vec{x} + dz^2)\,,
 \end{equation}
 where  $ (t,\vec{x})\in \mathcal{R}^{1,3} $ and $ z\in (0,\infty)$ is called radial coordinate.
For the action, we choose  
\begin{equation}
I \,=\, \frac{1}{2 g_5^ 2} \int d^4x dz \, \sqrt{-g} \,\,  \, \left\{  - \frac{1}{2 } F_{mn} F^{mn} - \mu^ 2(z)  V_m V^ m 
\,  \right\} \,\,, 
\label{vectorfieldaction}
\end{equation}
\noindent where $F_{mn} = \partial_mV_n - \partial_n V_m$ and the bulk mass has the form $ \mu(z)  = M^ 2 z^ 2 /R    $  where $ M$ is a mass parameter that plays the role of introducing the (heavy) quark mass in the model. In the conformal (AdS/CFT) case, where the mass $M$ is zero, the action (\ref{vectorfieldaction}) is gauge invariant and one can choose the gauge $V_z=0$. Then the remaining components $V_\mu$ of the vector field play the role of generators of   correlators  of the   boundary  currents $ J^\mu \,$. 
In the present case, there is no gauge invariance, but we assume that 
solutions of the vector field satisfying the condition $V_z=0$ work, as in the conformal case, as the sources of the current correlators.

The idea of using a bulk mass that varies with the radial coordinate as a type of infrared cut off in the 
gauge theory has the following interpretation:  the radial coordinate of AdS space is associated with 
the energy of the gauge theory.  A  bulk mass increasing  quadratically with $z$ implies that  low energies are represented by  bulk fields with  large mass. The limit of zero energy corresponds to an infinitely massive field.  So,  the bulk mass term suppress the low energies of the gauge theory.  

It is important to remark that the sign of the bulk mass term $ -  \mu^2 V_mV^m $ in eq. (\ref{vectorfieldaction}), with our metric convention,  is the opposite of the usual mass term of a vector field.
 That means, it  is like an imaginary mass term. 
 If one choose an action like  (\ref{vectorfieldaction}) but with an opposite mass term $ +  \mu^2 V_mV^m $
 it is known \cite{Afonin:2010hn,Afonin:2012xq}  that such a term can be eliminated by a field redefinition. 
This way, an action like (\ref{vectorfieldaction})  but with the a mass term $ +  \mu^2 V_mV^m $ 
can be maped into a soft wall action\cite{Karch:2006pv} without mass.
In the present case, this is not possible due to our choice of sign. 
Therefore, our action is not equivalent to a soft wall action. Thus, the new model that we are proposing for heavy vector mesons is not equivalent to a soft wall model.

The second energy parameter that is needed to represent the heavy vector mesons, corresponding to
 the quark anti-quark interaction,  is introduced in the model as the inverse of the position
  $ z = z_0$ of the radial AdS coordinate where the correlation functions of the gauge theory currents are  calculated. We take the prescription:
\begin{equation}
 \langle 0 \vert \, J_\mu (x) J_\nu (y) \,  \vert 0 \rangle \, =\, \frac{\delta}{\delta V^{0\mu}(x)} \frac{\delta}{\delta V^{0\nu}(y)}
 \exp  \left( - I_{on shell} \right)\,,
\end{equation}
\noindent where the source of the current operator is the value of the bulk field at the finite location $z_0$:  $ V^0_{\mu}(x) = \lim_{z\to z_0} V_\mu (x,z) $  and the on shell action is obtained by constraining the action  of eq.(\ref{vectorfieldaction}) to the AdS slice $ z_0 \le z < \infty $. This means the on shell action is:
 \begin{equation}
I_{on \, shell }\,=\, - \frac{1}{2 {\tilde g}_5^2}  \, \int d^4x \,\,\left[  \frac{ 1 }{z} V_\mu \partial_z V^ \mu 
\right]_{_{ \! z = z_0 }}^{z \to \infty}
 \,,
\label{onshellaction}
\end{equation}
where we introduced $ {\tilde g}_5^2 =  g_5^2 /R $, the relevant dimensionless coupling of the bulk vector field.  A similar calculation of two point functions at a non-vanishing position of the radial AdS coordinate
was discussed in refs. \cite{Evans:2006ea,Afonin:2011ff}, in the context of an AdS/QCD model
with an exponential dilaton background.  
 
One can represent the vector field $ V_\mu (x,z) $ in momentum space and use the  decomposition
 \begin{equation} 
 V_\mu (p,z) \,=\, v (p,z) V^0_\mu ( p ) \,,
 \label{Bulktoboundary}
\end{equation}  
 \noindent where    $ v (p,z)   $ is the bulk to boundary propagator that satisfies the equation of motion:
 \begin{equation}
\partial_z \Big( \frac{ 1 } { z}  \partial_z v (p,z) \Big) + \Big(  \frac{ - p^ 2 }{z} - M^ 4 z \Big)   v (p,z) \,=\, 0   \,.
\label{BulktoboundaryEOM}
\end{equation} 
The boundary condition:  
 \begin{equation} 
 \lim_{z\to z_0 }v (p, z) = 1\,,
 \label{boundarycond}
\end{equation} 
imposes that $ V^0_\mu ( p ) $ acts as the source of current-current correlators.
The upper limit of the on shell action in eq. (\ref{onshellaction}) is cancelled imposing 
 Neumann boundary condition at infinity: 
 \begin{equation}
 \lim_{z\to \infty }  \left( \frac{\partial v}{\partial z } \right)  = 0\,. 
 \label{BC2}
 \end{equation}

It is important to remark that when one uses a massless vector field to generate the correlators of the gauge current by means of a coupling term of the form: $ \int d^4x V^\mu J_\mu $, gauge invariance of the field 
implies conservation of the current. In the present case, where we use a massive vector field, the coupling to the currents does not guarantee by itself the conservation of the current. We will assume that the current is conserved, since it comes from a gauge invariant theory. The vector field is just as an external source that generates the expectation values of the currents. So, in momentum space the currents are transversal and the vacuum expectation value of product of currents has the structure:

 \begin{equation}  
   \int d^4x \,\, e^{-ip\cdot x} \langle 0 \vert \, J_\mu (x) J_\nu (0) \,  \vert 0 \rangle  \, =\,
 \left(  \eta_{\mu\nu} -  \frac {p_\mu p_\nu}{ p^2}    \right) \, G ( p^2 ) \,  .  
 \end{equation}

The AdS/CFT prescription, proposed in refs. \cite{Gubser:1998bc,Witten:1998qj},  is to take the on shell action as the generator of the gauge theory correlators.  Correspondingly, the bulk fields at $z=0$ play the role of the generators of the boundary theory correlators. 
Here an analogous prescription is used. The difference is that we take the bulk fields at the finite position $ z= z_0$ as the sources for the gauge theory correlators and assume that the same relations between bulk fields and boundary operators of the AdS/CFT correspondence are valid.   
 Correspondingly, the generator of correlation function is:
\begin{equation}
I_{on \, shell }\,=\, - \frac{1}{2 {\tilde g}_5^2}  \, \int d^4x \,\,\left[  \frac{ 1 }{z} V_\mu \partial_z V^ \mu 
\right]_{_{ \! z = z_0 }} 
 \,.
\label{onshellactionzero}
\end{equation}
This holographic prescription  provides an expression for the two point function in terms of 
the bulk to boundary propagator: 
\begin{equation}
 G ( p^2 )  \, =\,  -  \lim_{z\to z_0} \left(  \frac{1}{{\tilde g}_5^ 2 \, z  }   \frac{  \partial v (p,z)  }{ \partial  z  } \right) 
    \, .  
 \label{hol2point}
\end{equation} 
The equation of motion (\ref{BulktoboundaryEOM})  with Neumann boundary condition at infinity, has the solution:
\begin{equation}
 v (p,z) = C(p)  exp( - M^2 z^ 2 / 2) \, \,U (p^ 2/ 4M^ 2 , 0, M^2 z^ 2 )\,,
\end{equation}
where $U(a,b,x)$ is the confluent hypergeometric Kummer function as defined in \cite{NIST}  and  $C(p) $ is an 
arbitrary factor that does not depend on $z$.
Following a similar procedure as in  \cite{Afonin:2011ff},   one can build the bulk to boundary propagator, satisfying  $v (p,z_0 ) = 1$, by simply choosing $C (p) $ appropriately:
 \begin{equation} 
 v (p,z ) \, = \, \frac{ e^{- M^2 z^ 2 / 2} \, U (p^ 2/ 4M^ 2 , 0, M^2 z^ 2 ) }{  e^{- M^2 z_0^ 2 / 2}  \,  U (p^ 2/ 4M^ 2 , 0, M^2 z_0^ 2 )}\,.
 \label{bulktoboundary2}
\end{equation} 
Inserting the bulk to boundary propagator (\ref{bulktoboundary2})  in our holographic two point function of eq. (\ref{hol2point}) one finds:
 \begin{equation} 
G (p^2)  \, = \,\frac{1}{{\tilde g}_5^ 2 \,   }   \frac{ M^ 2 \, U ( p^ 2/ 4M^ 2 , 0, M^2 z_0^ 2 ) 
- \frac{p^ 2}{2}  U (1 + p^ 2/ 4M^ 2 , 0, M^2 z_0^ 2 ) }{    U ( p^ 2/ 4M^ 2 , 0, M^2 z_0^ 2 )}\,.
 \label{bulktoboundary3}
\end{equation}  
We associate the poles of $ G (p^2) $ with the masses of the states of the theory:  $ p_n^2 = -m_n^2 $.  

 At this point it is interesting to compare the approach developed here with the hard wall model  \cite{BoschiFilho:2002ta,BoschiFilho:2002vd}.  
In the hard wall case, the space ranges from $z =0$ to a maximum value $z = z_{_{HW}}$. The position $z=0$,  where one calculates the correlation functions, plays the role of an ultraviolet boundary corresponding to infinite energy. The position $z_{_{HW}}$ represents a hard infrared  cutoff (hard wall), where the field solutions corresponding to the states satisfy Dirichlet or Neuman boundary conditions. 
There is just one energy parameter in the hard wall model: $ 1/z_{_{HW}}$.

In contrast, the model developed here is defined in the region $ z_0 \le z  < \infty $. The position $z=z_0$ is a boundary where we calculate the correlation functions.   The masses are defined as the poles of eq.  (\ref{bulktoboundary3}) 
that come from the zeroes of the denominator of this equation. So, the solutions of the equation of motion, that have the form $ exp( - M^2 z^ 2 / 2) \, \,U (p^ 2/ 4M^ 2 , 0, M^2 z^ 2 )\,$  vanish at $ z = z_0$,  for $p^2 = -m_n^2 $.  Note that the normalization condition of eq. (\ref{boundarycond})  
 is achieved by writting $ v (p,z) = C(p)   exp( - M^2 z^ 2 / 2) \, \,U (p^ 2/ 4M^ 2 , 0, M^2 z^ 2 )\,$ and choosing $C(p) $ to be the inverse of the value of the function $exp( - M^2 z^ 2 / 2) \, \,U (p^ 2/ 4M^ 2 , 0, M^2 z^ 2 )\,$ at $z \to z_0$.
The fact that the solutions of the equation of motion corresponding to the physical states vanish at $z = z_0$
means that this position represents a hard wall. 
On the other hand,  the present model has also a smooth infrared cutoff represented by the mass term 
$ \mu(z)  = M^ 2 z^ 2 /R    .$ 
So, there are two energy parameters: $1/z_0 $ and $M$.

 In the next section we show the  mass spectra for charmonium and bottomonium S-wave states, obtained from the poles of eq. (\ref{bulktoboundary3}).

\section{ Model versus experimental data }


\begin{table}
\parbox{.45\linewidth}{
 \centering
\begin{tabular}[c]{|c|c|}
\hline 
\multicolumn{2}{|c|}{   Charmonium  Masses   } \\
\hline
  State & Mass (MeV)     \\
\hline
$ \,\,\,\, 1S  \,\,\,\,$ & $3096.916 \pm 0.011 $  \\ 
\hline
$ \,\,\,\,   2S  \,\,\,\,$ & $ 3686.109 \pm 0.012 $   \\
\hline 
$ \,\,\,\,3S \,\,\,\,$ & $ 4039 \pm 1 $  \\ 
\hline
\,\,\,\, $ 4S$  \,\,\,\,&$ 4421 \pm 4 $     \\
\hline
 \end{tabular}
 \caption{Experimental masses    for the Charmonium S-wave resonances  from \cite{Agashe:2014kda}.  }
 }
\hfill
\parbox{.45\linewidth}{       
 \centering
\begin{tabular}[c]{|c|c|}
\hline 
\multicolumn{2}{|c|}{  Bottomonium  Masses   } \\
\hline
State  &  Mass (MeV)    \\
\hline
$\,\,\,\, 1S \,\,\,\,$ & $ 9460.3\pm 0.26 $  \\ 
\hline
$\,\,\,\, 2S \,\,\,\,$ & $ 10023.26 \pm 0.32 $   \\
\hline 
$\,\,\,\,3S \,\,\,\,$ & $ 10355.2 \pm 0.5 $   \\ 
\hline
$ \,\,\,\, 4S  \,\,\,\,$ & $ 10579.4 \pm 1.2 $   \\
\hline
$ \,\,\,\, 5S  \,\,\,\,$ & $ 10860 \pm 11 $   \\
\hline
$ \,\,\,\, 6S  \,\,\,\,$ & $ 11019 \pm 8 $   \\
\hline
\end{tabular}   
\caption{Experimental masses  for the Bottomonium S-wave resonances  from \cite{Agashe:2014kda}.  }
 }
\end{table}
 

The experimental values for the masses of  S-wave states  of charmonium and bottomonium from Particle Data Group Collaboration \cite{Agashe:2014kda} are show on
tables 1 and 2, with the corresponding uncertainties.   The best fit of the model for the masses of the heavy vector mesons  is obtained for the choice of parameters:
$$ M_c = 0.74 {\textrm GeV }  ; \,\, M_b = 1.35  {\textrm GeV }  ; \,\, 1/ z_0 = 0.25 {\textrm GeV },  $$
where $  M_c $ and $ M_b $ are the values of the parameter $ M$  of the model used for the cases of charmonium and bottomonium respectively. The energy parameter $1/ z_0$ represents the energy levels of the interaction between the quark and the anti-quark, that is expected to be dominated by color interaction, so for consistency we use
the same value for the two flavors of vector mesons.

\begin{table}
\parbox{.45\linewidth}{
\centering
\begin{tabular}[c]{|c|c|}
\hline 
\multicolumn{2}{|c|}{   Charmonium  Results   } \\
\hline
  State & Mass (MeV)     \\
\hline
$ \,\,\,\, 1S  \,\,\,\,$ & $ 3075.5 \,\, (0.68 \%) $  \\ 
\hline
$ \,\,\,\,   2S  \,\,\,\,$ & $  3664.5  \,\, (0.58 \%) $   \\
\hline 
$ \,\,\,\,3S \,\,\,\,$ & $  4118.2    \,\, (1.20 \% ) $  \\ 
\hline
\,\,\,\, $ 4S$  \,\,\,\,&$  4502.5  \,\, (1.84 \% )  $     \\
\hline
 \end{tabular}
\caption{Masses of charmonium S-wave resonances from the holographic model with 
 $ M_c = 0.74 $  GeV  and $1/ z_0 = 0.25 $ GeV. The percentages are the deviations with respect to average experimental values.}
}
\hfill
\parbox{.45\linewidth}{
\centering
\begin{tabular}[c]{|c|c|}
\hline 
\multicolumn{2}{|c|}{  Bottomonium  Results    } \\
\hline
State  &  Mass (MeV)    \\
\hline
$\,\,\,\, 1S \,\,\,\,$ & $  8662.37 \,\,(8.43 \%)  $  \\ 
\hline
$\,\,\,\, 2S \,\,\,\,$ & $ 9625.72 \,\,(3.96 \%)  $   \\
\hline 
$\,\,\,\,3S \,\,\,\,$ & $ 10383.5 \,\,( 0.27 \%)  $   \\ 
\hline
$ \,\,\,\, 4S  \,\,\,\,$ & $ 11033.6 \,\,(4.28 \%)  $   \\
\hline
$ \,\,\,\, 5S  \,\,\,\,$ & $ 11613.7 \,\,(6.94  \%) $   \\
\hline
$ \,\,\,\, 6S  \,\,\,\,$ & $ 12143.2 \,\,(10.2  \%)  $   \\
\hline
\end{tabular}   
\caption{Masses of bottomonium S-wave resonances from the holographic model with 
 $ M_b = 1.35    $  GeV  and $1/ z_0 = 0.25 $  GeV. The percentages are the deviations with respect to average experimental values. }
}
\end{table}

We show in tables 3 and 4 the results of the holographic model, with the percentage deviations with respect to (average) experimental data.   
 As a measure of the predictability of the model, one can define the rms error for estimating $N$ quantities using a model with $N_p$ parameters as:
 \begin{equation}
 \delta_{rms} = \sqrt{ \frac{1}{(N - N_ p )}\sum_i^N  \left( \frac{\delta O_i}{O_i} \right)^ 2 }\,,
 \label{error}
 \end{equation}
 \noindent where $O_i$ is the average experimental value and $\delta O_i$ is the deviation of the value given 
 by the model.    We find for our estimate of 10 states with 3 parameters:  $  \delta_{rms}  =  6.1  \, \% $. 
 In particular, considering separately the charmonium states, we have an rms error of $  \delta^c_{rms}  =  1.7 \, \% $.
   
 It is interesting to mention that  although the experimental data available for charmonium at present time are conclusive only for the masses of the first four states: 1S - 4S,  there is also some  indication from experimental data on  $e^+ e^- $ anihilation analyzed by BaBar collaboration about higher S-wave states. 
In ref.   \cite{vanBeveren:2010jz} masses for   5S  up to  8S states are estimated, based on data from BaBar published in ref. \cite{Aubert:2009aq}. 
 We present on table 5 a comparison of the masses from this reference and the results of the holographic model 
 for these states, where again a remarkable agreement is found.

\begin{table}[h]
\centering
\begin{tabular}[c]{|c|c|c|}
\hline 
\multicolumn{3}{|c|}{   Charmonium higher excitations   } \\
\hline
  State &  Possible Mass  (MeV) \cite{vanBeveren:2010jz} & Holographic result      \\
\hline
$ \,\,\,\, 5S  \,\,\,\,$ & $  4780 $ & 4842  \,\,(1.30 \%) \\ 
\hline
$ \,\,\,\,   6S  \,\,\,\,$ & $ 5090  $ & 5150  \,\,(1.18  \%) \\
\hline 
$ \,\,\,\,7S \,\,\,\,$ & $ 5440  $ & 5434.4  \,\,(0.1  \%)  \\ 
\hline
\,\,\,\, $ 8S$  \,\,\,\,&$ 5910 $  &  5699.3  \,\,(3.56  \%)   \\
\hline
 \end{tabular}
 \caption{ Possible higher S-wave charmonium states compared with the results of the holographic model using $ M_c = 0.74 $  GeV  and $1/ z_0 = 0.25 $  GeV.  
 The percentages in parenthesis are  the deviations of the model with respect to estimates from \cite{vanBeveren:2010jz}.
  }
\end{table}

\section{  Final Comments}
 
 The results of the tables presented in the previous section and the rms errors of  $ 6.1 \, \%$ show that the model proposed here 
 is indeed  capturing the behavior of the mass spectra of heavy vector meson radial excitations.  
The large energy scale, related to the quark mass, was introduced as a varying  bulk mass while the small scale associated with quark anti-quark interaction showed up from the  position  of anti-de Sitter space where operator expectation values are calculated.

For completeness, we mention that the calculation of correlation functions at a finite position in AdS space 
appeared, in the context of  exponential (soft wall) dilaton background, in  \cite{Afonin:2011ff} and was used recently in  \cite{Braga:2015jca}  to describe the observed  behavior of decay constants of vector mesons. 

As we have shown here,  it is the combination of the varying  bulk mass with the  definition 
of the gauge theory correlators at a finite location of AdS space that  provides the appropriate description of  heavy vector mesons masses. 

\bigskip
  
\noindent {\bf Acknowledgments:}  N.B. and S.D. are partially supported by CNPq and M.A.M. is supported by Vicerrectoria de Investigaciones de La Universidad
de los Andes.

 \end{document}